\documentclass[12pt]{article}

\usepackage{epsfig}

\newlength{\dinwidth} \newlength{\dinmargin}
\def\beq{\begin{equation}}
\def\eeq{\end{equation}}
\def\beqa{\begin{eqnarray}}
\def\eeqa{\end{eqnarray}}

\begin{document}

\begin{center}
{\Large \bf Renormalization group evolution of collinear and infrared divergences}
\end{center}
\vspace{1mm}
\begin{center}
{\large Nikolaos Kidonakis}\\
\vspace{2mm}
{\it Kennesaw State University,  Physics \#1202, Kennesaw, GA 30144, USA}
\end{center}

\begin{abstract}
I discuss collinear and infrared divergences in QCD cross sections 
with massless and massive final-state particles. I present the two-loop 
renormalization group evolution and resummation in terms of anomalous 
dimensions, and I show specific results for a variety of QCD hard-scattering 
processes.
\end{abstract}

\vspace{3mm}

\begin{center}
{\large \bf RESUMMATION OF COLLINEAR AND SOFT CORRECTIONS}
\end{center}

Soft-gluon corrections arise in scattering cross sections from incomplete 
cancellations of infrared divergences in virtual diagrams 
and real diagrams with low-energy (soft) gluons.
At $n$th order in the perturbative series, these soft corrections are of 
the form  $[(\ln^k(s_4/M^2))/s_4]_+$ with $M$ a hard scale, 
$k \le 2n-1$ and $s_4$ the kinematical distance from threshold.
The leading (double) logarithms arise from collinear and soft radiation. 
Also purely collinear terms $(1/M^2) \ln^k(s_4/M^2)$
appear in the cross section. 

Soft-gluon corrections are dominant near threshold and they can be shown to
exponentiate, so these corrections can be resummed.
Resummation follows from factorization properties of the cross section and 
renormalization group evolution (RGE) \cite{CLS,NKGS} (for further recent 
studies see Refs. [3-14]). 
At next-to-leading-logarithm (NLL) accuracy this requires one-loop 
calculations in the eikonal approximation. Recently results have been derived 
at next-to-NLL (NNLL), with the completion
of two-loop calculations for soft anomalous dimensions for processes 
with massless and massive partons in various approaches [3,6-14]. 
Approximate NNLO and higher-order cross sections have also been derived from 
the expansion of the resummed cross sections.

The cross section factorizes as
$\sigma=(\prod \psi) \; H_{IL} \, S_{LI} \; (\prod J)$, where $\psi$ are
functions for the incoming partons, $J$ are final-state jet functions, 
$H$ is the hard-scattering function, and $S$ is the soft-gluon function 
describing noncollinear soft-gluon emission \cite{NKGS}.
We use RGE to evolve the $S$ function associated with soft-gluon emission 
$$
\left(\mu {\partial \over \partial \mu}
+\beta(g_s){\partial \over \partial g_s}\right)\,S_{LI}
=-(\Gamma^\dagger_S)_{LB}S_{BI}-S_{LA}(\Gamma_S)_{AI}
\nonumber
$$
where $\Gamma_S$ is the soft anomalous dimension, a matrix in 
color space and a function of the kinematical invariants of the 
process \cite{NKGS}.

Solving the RGE for the soft function and the other functions in the factorized
cross section, we find the following result for the resummed cross section
in Mellin moment space, with $N$ the moment variable, 
\beqa
{\hat{\sigma}}^{res}(N) &=&
\exp\left[ \sum_i E_i(N_i)\right] \, \exp\left[ \sum_j E'_j(N')\right]\;
\exp \left[\sum_i 2 \int_{\mu_F}^{\sqrt{s}} \frac{d\mu}{\mu}\;
\gamma_{i/i}\left({\tilde N}_i, \mu\right)\right] \;
\nonumber\\ && \hspace{-15mm} \times 
{\rm tr} \left\{H
\exp \left[\int_{\sqrt{s}}^{{\sqrt{s}}/{\tilde N'}}
\frac{d\mu}{\mu}
\Gamma_S^{\dagger}(\mu)\right] \;
S(\sqrt{s}/\tilde N') 
\exp \left[\int_{\sqrt{s}}^{{\sqrt{s}}/{\tilde N'}}
\frac{d\mu}{\mu} \Gamma_S(\mu)\right] \right\} \,.
\nonumber 
\eeqa

Collinear and soft radiation from the incoming partons is resummed in the 
exponent
$$
E_i(N_i)=
\int^1_0 dz \frac{z^{N_i-1}-1}{1-z}\;
\left \{\int_1^{(1-z)^2} \frac{d\lambda}{\lambda}
A_i\left(\lambda s\right)
+D_i\left[(1-z)^2 s\right]\right\} .
$$
Purely collinear terms can be derived by replacing $\frac{z^{N-1}-1}{1-z}\;$ 
by $\;- z^{N-1}$ above.

Collinear and soft radiation from outgoing massless quarks and gluons 
is resummed in the second exponent 
$$
E'_j(N')=
\int^1_0 dz \frac{z^{N'-1}-1}{1-z}
\left \{\int^{1-z}_{(1-z)^2} \frac{d\lambda}{\lambda}
A_i \left(\lambda s\right)
+B_i\left[(1-z)s\right]
+D_i\left[(1-z)^2 s\right]\right\} .
$$
The quantities $A$, $B$, and $D$ have well-known perturbative expansions in 
$\alpha_s$. 
The factorization scale, $\mu_F$, dependence in the third exponent is 
controlled by parton anomalous dimensions 
$\gamma_{i/i}=-A_i \ln {\tilde N}_i +\gamma_i$. 
Noncollinear soft gluon emission is controlled by the 
process-dependent soft anomalous dimension $\Gamma_S$. 

We determine $\Gamma_S$ from the coefficients of ultraviolet poles in 
dimensionally regularized eikonal diagrams [2,6,11-15]. 
We perform the calculations in momentum space and Feynman gauge.
Complete two-loop results have been derived for the 
soft anomalous dimensions for 
$e^+ e^- \rightarrow t {\bar t}$ \cite{NK2l}, 
$t{\bar t}$ hadroproduction \cite{NKttbar},
$t$-channel \cite{NKtch} and $s$-channel \cite{NKsch} single top production,  
$t W^-$ and  $t H^-$ production \cite{NKtWH}, and 
direct photon and $W$ production at large $Q_T$.
We write the perturbative series for the soft anomalous dimension
$\Gamma_S=(\alpha_s/\pi) \Gamma_S^{(1)}+(\alpha_s/\pi)^2 \Gamma_S^{(2)}+\cdots$
and determine $\Gamma_S^{(1)}$ and $\Gamma_S^{(2)}$ for these processes.

\vspace{3mm} 

\begin{center}
{\large \bf TWO-LOOP SOFT ANOMALOUS DIMENSIONS}
\end{center}
\begin{center}
{\large \bf Top-antitop production in hadron colliders}
\end{center}

The soft anomalous dimension matrix for the partonic process 
$q{\bar q} \rightarrow t{\bar t}$ is 
a $2 \times 2$ matrix \cite{NKGS,NKttbar} 
\beqa
\Gamma_{S\, q{\bar q}}=\left[\begin{array}{cc}
\Gamma_{q{\bar q} \, 11} & \Gamma_{q{\bar q} \, 12} \\
\Gamma_{q{\bar q} \, 21} & \Gamma_{q{\bar q} \, 22}
\end{array}
\right] .
\nonumber
\eeqa
At one loop, in a singlet-octet color basis, we find 
\beqa
&& \Gamma_{q{\bar q} \,11}^{(1)}=-C_F \, [L_{\beta}+1] 
\hspace{15mm}
\Gamma_{q{\bar q} \,21}^{(1)}=
2\ln\left(\frac{u_1}{t_1}\right) \hspace{15mm}
\Gamma_{q{\bar q} \,12}^{(1)}=
\frac{C_F}{C_A} \ln\left(\frac{u_1}{t_1}\right) 
\nonumber \\ &&
\Gamma_{q{\bar q} \,22}^{(1)}=C_F
\left[4\ln\left(\frac{u_1}{t_1}\right)
-L_{\beta}-1\right]
+\frac{C_A}{2}\left[-3\ln\left(\frac{u_1}{t_1}\right)
+\ln\left(\frac{t_1u_1}{s m^2}\right)+L_{\beta}\right]
\nonumber
\eeqa
where 
$L_{\beta}=[(1+\beta^2)/(2\beta)]\ln[(1-\beta)/(1+\beta)]$
with $\beta=\sqrt{1-4m^2/s}$ and $m$ the top quark mass.

At two loops, we find \cite{NKttbar}
\beqa
&& \Gamma_{q{\bar q} \,11}^{(2)}=\frac{K}{2} \Gamma_{q{\bar q} \,11}^{(1)}
+C_F C_A \, M_{\beta} \hspace{20mm}
\Gamma_{q{\bar q} \,22}^{(2)}=
\frac{K}{2} \Gamma_{q{\bar q} \,22}^{(1)}
+C_A\left(C_F-\frac{C_A}{2}\right) \, M_{\beta} 
\nonumber \\ &&
\Gamma_{q{\bar q} \,21}^{(2)}=
\frac{K}{2}  \Gamma_{q{\bar q} \,21}^{(1)} +C_A N_{\beta} \ln\left(\frac{u_1}{t_1}\right) \hspace{15mm} 
\Gamma_{q{\bar q} \,12}^{(2)}=
\frac{K}{2} \Gamma_{q{\bar q} \,12}^{(1)} -\frac{C_F}{2} N_{\beta} \ln\left(\frac{u_1}{t_1}\right) 
\nonumber
\eeqa
where $K$ is a two-loop constant, $M_{\beta}$ is a part of the two-loop cusp 
anomalous dimension \cite{NK2l}, and $N_{\beta}$ is a subset of the terms 
of $M_{\beta}$.

Similar results have been derived for the $gg\rightarrow t{\bar t}$ channel \cite{NKttbar}.

\vspace{2mm}

\begin{center}
{\large \bf Single top quark production} 
\end{center}

We begin with the soft anomalous dimension for $t$-channel single top 
production \cite{NKtch}. Here we show results only for the 11 element of the matrix.
At one loop 
$$
{\Gamma}_{{\rm t\mbox{-}ch}\, 11}^{(1)}
=C_F \left[\ln\left(\frac{-t}{s}\right)
+\ln\left(\frac{m^2-t}{m\sqrt{s}}\right)-\frac{1}{2}\right] .
$$
At two loops \cite{NKtch}
$$
\Gamma_{{\rm t\mbox{-}ch}\, 11}^{(2)}=\frac{K}{2}
\Gamma_{{\rm t\mbox{-}ch}\, 11}^{(1)}
+C_F C_A \frac{(1-\zeta_3)}{4} .
$$

We continue with the soft anomalous dimension for $s$-channel 
single top production \cite{NKsch}, again showing only the 11 matrix element:
$$
\Gamma_{{\rm s\mbox{-}ch}\, 11}^{(1)}=C_F 
\left[\ln\left(\frac{s-m^2}{m\sqrt{s}}\right)
-\frac{1}{2}\right]\, , \quad \quad
\Gamma_{{\rm s\mbox{-}ch}\, 11}^{(2)}=\frac{K}{2} 
\Gamma_{{\rm s\mbox{-}ch}\, 11}^{(1)}
+C_F C_A \frac{(1-\zeta_3)}{4} .
$$

Finally we present the soft anomalous dimension for the associated production 
of a top quark with a $W^-$ or $H^-$. 
Relevant two-loop eikonal diagrams are shown in Fig. 1
(there are also additional top-quark self-energy graphs).

\begin{figure}
\begin{center}
  \includegraphics[height=.25\textheight]{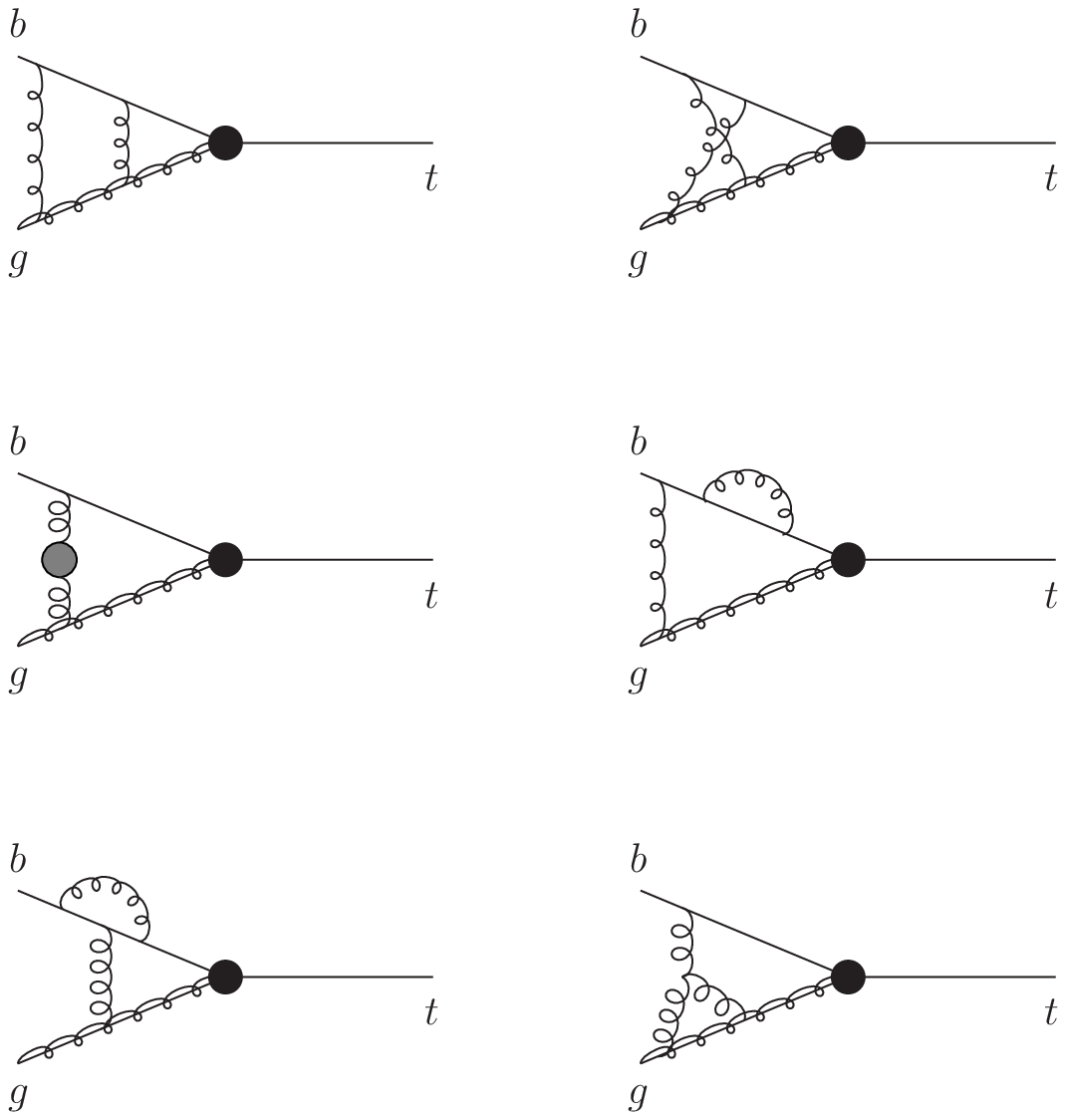}
\hspace{3mm}
  \includegraphics[height=.25\textheight]{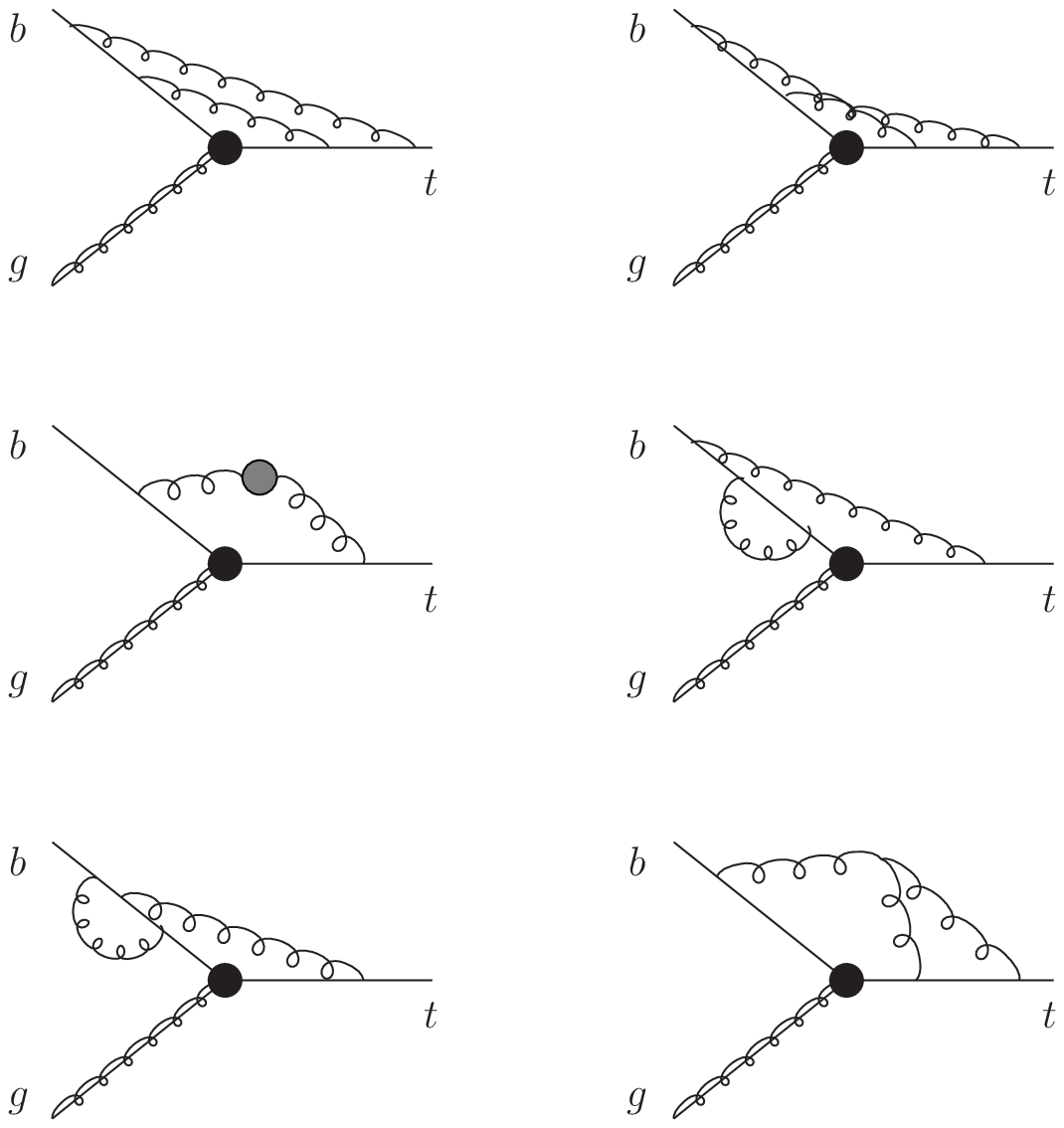}
\hspace{3mm}
  \includegraphics[height=.25\textheight]{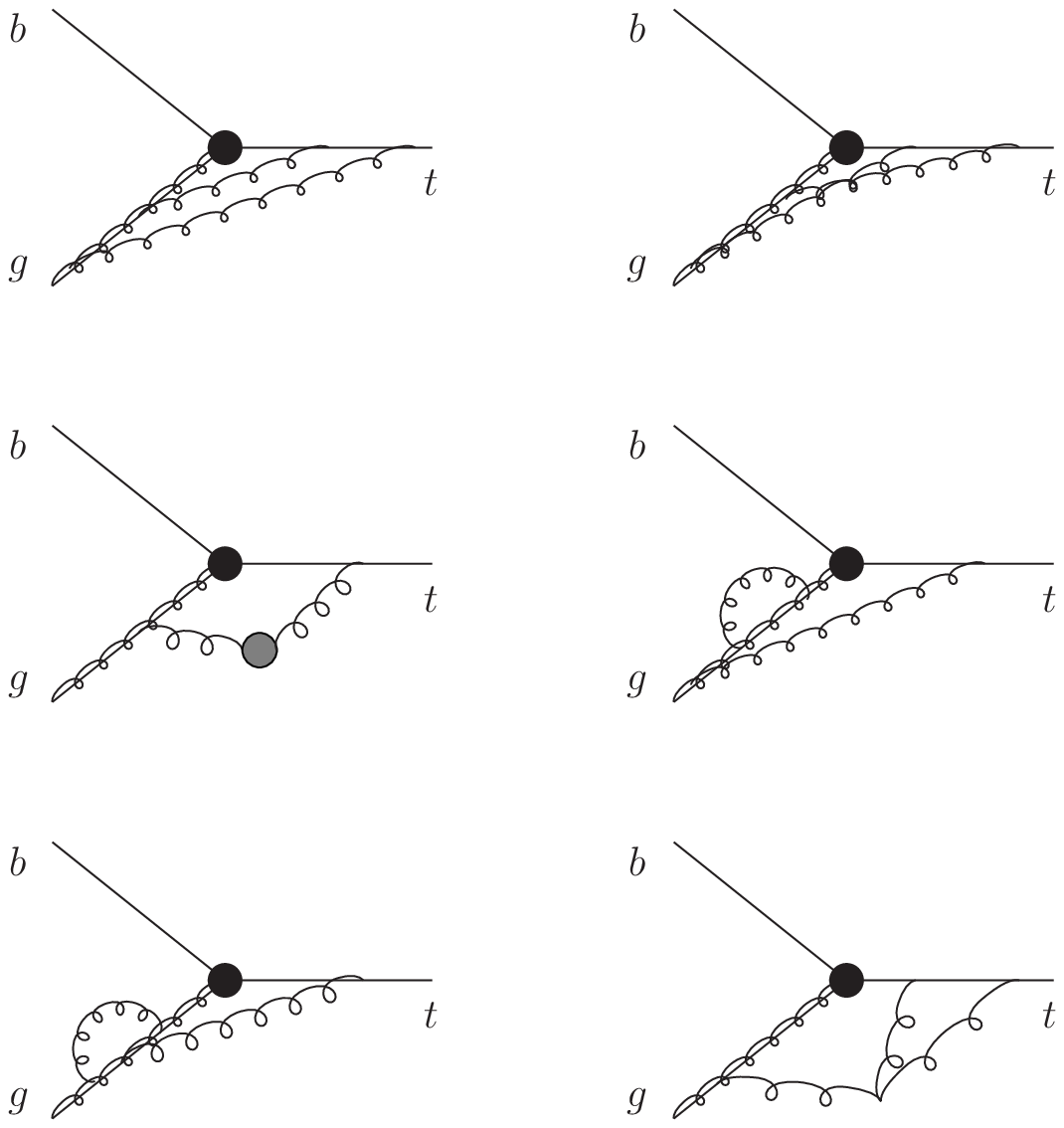}
  \caption{Two-loop eikonal diagrams for $tW$ production.}
\end{center}
\end{figure}

The soft anomalous dimension for $bg \rightarrow tW^-$ 
(or $bg \rightarrow tH^-$) is \cite{NKtWH}
$$
\Gamma_{S,\, tW^-}^{(1)}=C_F \left[\ln\left(\frac{m^2-t}{m\sqrt{s}}\right)
-\frac{1}{2}\right] +\frac{C_A}{2} \ln\left(\frac{m^2-u}{m^2-t}\right)
$$
$$
\Gamma_{S,\, tW^-}^{(2)}=\frac{K}{2} \Gamma_{S,\, tW^-}^{(1)}
+C_F C_A \frac{(1-\zeta_3)}{4} \, .
$$

\vspace{2mm}

\begin{center}
{\large \bf $W$-boson and direct photon production at large $p_T$}
\end{center}

One-loop results for the soft anomalous dimensions for $W$ (same as for direct photon) production 
have been known from \cite{NKVDD}. Here we also present new two-loop results.
 
For the process $qg\rightarrow Wq$ (or $qg\rightarrow \gamma q$)
the soft anomalous dimension is
$$
\Gamma_{S,\, qg\rightarrow Wq}^{(1)}=C_F \ln\left(\frac{-u}{s}\right)
+\frac{C_A}{2} \ln\left(\frac{t}{u}\right) \, , \quad \quad
\Gamma_{S,\, qg \rightarrow Wq}^{(2)}=\frac{K}{2} \Gamma_{S,\, qg \rightarrow Wq}^{(1)} \, .
$$

For the process $q {\bar q} \rightarrow Wg$ 
(or $q {\bar q}\rightarrow \gamma g$) the soft anomalous dimension is
$$
\Gamma_{S,\, q{\bar q}\rightarrow Wg}^{(1)}=\frac{C_A}{2} \ln\left(\frac{tu}{s^2}\right) \, , \quad \quad
\Gamma_{S,\, q{\bar q} \rightarrow Wg}^{(2)}=\frac{K}{2} \Gamma_{S,\, q{\bar q} \rightarrow Wg}^{(1)} \, .
$$

\vspace{3mm}

\begin{center}
{\large \bf ACKNOWLEDGMENTS}
\end{center}

This work was supported by the National Science Foundation under 
Grant No. PHY 0855421.


\vspace{3mm}

\begin{thebibliography}{9}

\bibitem{CLS}
H. Contopanagos, E. Laenen, and G. Sterman, \emph{Nucl. Phys. B} \textbf{484},
303 (1997) [hep-ph/9604313].

\bibitem{NKGS}
N. Kidonakis and G. Sterman, \emph{Nucl. Phys. B} \textbf{505}, 321 (1997) [hep-ph/9705234].

\bibitem{ADS}
S.M. Aybat, L.J. Dixon, and G. Sterman, \emph{Phys. Rev. D} \textbf{74}, 074004 (2006) [hep-ph/0607309].

\bibitem{DMS}
L.J. Dixon, L. Magnea, and G. Sterman, \emph{JHEP} \textbf{08}, 022 (2008) 
[arXiv:0805.3515 [hep-ph]].

\bibitem{GM}
E. Gardi and L. Magnea, \emph{JHEP} \textbf{03}, 079 (2009) [arXiv:0901.1091 [hep-ph]].

\bibitem{NK2l}
N. Kidonakis, \emph{Phys. Rev. Lett.} \textbf{102}, 232003 (2009)  
[arXiv:0903.2561 [hep-ph]].

\bibitem{MSS} 
A. Mitov, G. Sterman, and I. Sung, \emph{Phys. Rev. D} \textbf{79}, 094015 (2009) 
[arXiv:0903.3241 [hep-ph]]. 

\bibitem{BNm} 
T. Becher and M. Neubert, \emph{Phys. Rev. D} \textbf{79}, 125004 (2009)  
[arXiv:0904.1021 [hep-ph]].

\bibitem{BFS} 
M. Beneke, P. Falgari, and C. Schwinn, \emph{Nucl. Phys. B} \textbf{828}, 69 (2010)  
[arXiv:0907.1443 [hep-ph]].

\bibitem{FNPY} 
A. Ferroglia, M. Neubert, B. Pecjak, and L. Yang, 
\emph{JHEP} \textbf{11}, 062 (2009) [arXiv:0908.3676 [hep-ph]]. 

\bibitem{NKsch}
N. Kidonakis, \emph{Phys. Rev. D} \textbf{81}, 054028 (2010)  
[arXiv:1001.5034 [hep-ph]]. 

\bibitem{NKtWH}
N. Kidonakis, \emph{Phys. Rev. D} \textbf{82}, 054018 (2010) 
[arXiv:1005.4451 [hep-ph]].

\bibitem{NKttbar}
N. Kidonakis, \emph{Phys. Rev. D}  \textbf{82}, 114030 (2010)   
[arXiv:1009.4935 [hep-ph]].

\bibitem{NKtch}
N. Kidonakis, \emph{Phys. Rev. D} \textbf{83}, 091503(R) (2011) 
[arXiv:1103.2792 [hep-ph]].

\bibitem{NKVDD}
N. Kidonakis and V. Del Duca, \emph{Phys. Lett. B} \textbf{480}, 87 (2000) [hep-ph/9911460].

\end{thebibliography}
\end{document}